\documentclass[twocolumn,superscriptaddress]{revtex4}
\usepackage{graphicx}
\usepackage{epsfig}
\usepackage{amsmath}
\usepackage{amsfonts,amsbsy}
\usepackage{amssymb}
\usepackage{hyperref}
\usepackage{bm}
\usepackage{color}
\usepackage{enumerate}
\usepackage{verbatim}

\def\be{\begin{equation}}
\def\ee{\end{equation}}
\def\bea{\begin{eqnarray}}
\def\eea{\end{eqnarray}}

\begin{document}
\title{Relaxation dynamics and the free energy near the phase boundary of the 3D kinetic Ising model}
\author{Ranran Guo}
\affiliation{Key Laboratory of Quark and Lepton Physics (MOE) and Institute of Particle Physics, Central China Normal University, Wuhan 430079, China}
\author{Xiaobing Li}
\affiliation{School of Physics and Electronic Engineering, Hubei University of Arts and Science, Xiangyang 441053, China}
\author{Yuming Zhong}
\affiliation{Key Laboratory of Quark and Lepton Physics (MOE) and Institute of Particle Physics, Central China Normal University, Wuhan 430079, China}
\author{Mingmei Xu}
\email{xumm@ccnu.edu.cn}
\affiliation{Key Laboratory of Quark and Lepton Physics (MOE) and Institute of Particle Physics, Central China Normal University, Wuhan 430079, China}
\author{Jinghua Fu}
\affiliation{Key Laboratory of Quark and Lepton Physics (MOE) and Institute of Particle Physics, Central China Normal University, Wuhan 430079, China}
\author{Yuanfang Wu}
\email{wuyf@ccnu.edu.cn}
\affiliation{Key Laboratory of Quark and Lepton Physics (MOE) and Institute of Particle Physics, Central China Normal University, Wuhan 430079, China}
\date{\today}
\begin{abstract}
We investigate relaxation dynamics along the entire first-order phase transition line by analyzing the time evolution of the free energy landscape in the three-dimensional kinetic Ising model. Near the critical temperature $T_{\rm c}$, the free energy structure is consistent with predictions from Landau-Ginzburg theory. At temperatures far below $T_{\rm c}$, however, fine structures in pre-equilibrium configurations  trap random initial states, causing a pronounced delay in equilibration - an effect we identify as ultra-slow relaxation. This phenomenon is characterized by a self-divergence of the relative variance of equilibration times, which we propose as a previously unrecognized hallmark of first-order phase transitions.
\end{abstract}

\maketitle

\section{Introduction}

The relaxation dynamics near phase boundaries is a fundamental question in non-equilibrium physics, with profound implications for material science, high energy nuclear physics, etc. While critical slowing down is well characterized through dynamical universality classes~\cite{1977}, the relaxation at first-order phase transitions (1st-PT) lacks complete theoretical and experimental understanding. 


The free energy as a primary thermodynamic potential exhibits intrinsic connections to relaxation dynamics~\cite{book1,1979-Ikeda,2001-Joo,2010-Fischer,1992-Berg,1995-Lee}. The connections manifest in several aspects: first, the free energy distinguishes stable and metastable states, guiding the direction of non-equilibrium evolution; second, the partial derivative of the free energy with respect to the order parameter determines near-equilibrium relaxation rates~\cite{book1}, governing the timescale of non-equilibrium evolution. Particularly at 1st-PT, the connection to multiple timescales has been studied. 


Earlier work shows that the relaxation time diverges near the spinodal point, termed as ``pseudo-critical slowing down"~\cite{1979-Ikeda}. This stems from the vanishing second derivative of free energy with respect to the order parameter which determines the inverse of the relaxation time. Numerical studies of Mott-Hubbard transition confirm this, i.e. the convergence rate of the solutions of the dynamical mean-field theory slows down significantly at two spinodal lines below $T_{\rm c}$~\cite{2001-Joo}.

Later, the tunneling time at 1st-PT is determined by the free energy barrier. For a two-phase coexistence system in a square of side length $L$, the interface length is approximately $2L$, creating a barrier $\Delta F\sim 2\sigma L$ (where $\sigma$ is line tension). Since $\tau^{\rm tunneling}\sim\exp(\Delta F)$, the tunneling time from one phase to the other grows exponentially with $L$, termed as ``exponential slowing down"~\cite{2010-Fischer,1992-Berg}. 


Similarly, the lifetime of the metastable state, defined by the relaxation time of a metastable phase towards the stable phase, is also determined by the free energy barrier, i.e. $\tau^{\rm lifetime}\sim\exp(2\beta\sigma L^{d-1})$~\cite{1995-Lee}, where $\beta$ is the inverse temperature and $d$ is the dimension of the system. The lifetime of the metastable state exhibits a similar dependence on system size as the tunneling time because the evolution from metastable to stable equilibrium states inherently involves tunneling through the free energy barrier.

In our recent work~\cite{XBLi1,XBLi2}, we introduced the equilibration time $\tau_{\rm eq}$, defined as the evolution time required for an initial configuration to reach equilibrium. We found that $\tau_{\rm eq}$ diverges at temperatures far below $T_{\rm c}$, a phenomenon we refered to as ``ultra-slow relaxation". Furthermore, we observed that $\tau_{\rm eq}$ scales with system size according to a power law, i.e. $\tau_{\rm eq}\sim L^{z}$, where the dynamic exponent $z$ increases as the temperature decreases. 


These findings indicate that 1st-PT involves more complex relaxation dynamics than continuous transitions. The dynamical slowing down observed across different timescales can be attributed to the structure of the underlying free energy landscape.



The exact calculation of free energy remains challenging even for simple systems like the 2D Ising model~\cite{complex}. Near critical points (CP), the Landau-Ginzburg theory provides a phenomenological framework by expanding the free energy in terms of a small order parameter, constrained by symmetry considerations. At zero field, this 
yields distinct landscape: a single minimum above $T_{\rm c}$, a flat-bottomed minimum at $T_{\rm c}$ and two degenerate minima below $T_{\rm c}$. These results are valid only near $T_{\rm c}$.  Applying a field breaks this degeneracy, creating stable and metastable states. 

Then, what would be the free energy landscape near the first-order phase boundary far below $T_{\rm c}$? Would it resemble the behavior near $T_{\rm c}$? In this paper, the Monte Carlo simulations of three-dimensional kinetic Ising model are performed by using the Metropolis algorithm. The fine structures and time evolution of the free energy landscape will be presented along and near the entire first-order phase transition line. These will give a good understanding on the ultra-slow relaxation at 1st-PT.


In this paper, we further examine the self-averaging properties of the equilibration time at the phase boundary. As we know, far from the phase boundary, the relative variance of thermodynamic quantities, i.e. the ratio of the variance to the square of the mean value, decays to zero with increasing system size - a phenomenon termed as self-averaging. This self-averaging property, however, breaks down at the CP, where the relative variance of magnetization or susceptibility approaches a constant value as the system size increases, indicating non-self-averaging behavior~\cite{nsa-cp-1,nsa-cp-2,nsa-cp-3,nsa-cp-4}. Non-self-averaging behavior at the 1st-PT has also been observed in the isotropic-to-nematic transition in liquid crystals~\cite{nsa-fo}. 

We compute the relative variance of the equilibration time $\tau_{\rm eq}$ and find that it approaches a constant value at $T_{\rm c}$ as the system size increases, indicating non–self-averaging behavior. In contrast, at 1st-PT, the relative variance of $\tau_{\rm eq}$ grows with system size, revealing a divergent behavior that differs from that at the critical point.


The paper is organized as follows: Section II introduces the kinetic Ising model and computational methods for free energy. Section III presents the free energy obtained by MC simulations and its time evolution. Section IV shows the relative variance of the equilibration time and a discussion on non-self-averaging behavior. Section V provides a summary and discussions. 

\section{Model and method}

The three-dimensional Ising model considers a simple cubic lattice composed of $N=L^3$ spins, where $L$ is called the system size. The microstate of the system can be represented by a series of spins, i.e. $\pmb{k}=\lbrace s_{i}\rbrace$, $s_{i}=\pm1$, $i=1,2,\cdots,N$. The total energy of the microstate $\pmb{k}$ with a constant nearest-neighbor interaction $J$ in a uniform external field $H$ is 
\begin{equation}
E_{\pmb{k}}=E_{\lbrace s_{i}\rbrace}=-J\sum_{\langle ij\rangle}s_{i}s_{j}-H\sum_{i=1}^N s_{i}.
\end{equation}
The first summation runs over all nearest-neighbor pairs. We adopt the conventional setting where $J$ (interaction strength) and $k_{\rm B}$ (Boltzman's constant) are both set to 1.


The magnetization per spin of the microstate $\pmb{k}$ reads
\begin{equation}
m_{\pmb{k}}=\frac{\sum_{i=1}^{N}s_i}{N}.
\end{equation}
An average of $m_{\pmb{k}}$ over all possible microstates serves as the order parameter. The first-order phase transition line lies at $H=0$ and $T<T_{\rm c}$ where $T_{\rm c}=4.51$~\cite{kc}.

In the kinetic Ising model~\cite{kinetic-Ising}, the Metropolis algorithm implements the Markov process via single-spin flips with acceptance ratio:
\begin{equation}
A({\pmb k}\rightarrow {\pmb k'})=\left\{\begin{array}{ll}
{\rm e}^{-(E_{\pmb k'}-E_{\pmb k})/k_{\rm B}T}&\text{if $E_{\pmb k'}-E_{\pmb k}>0$},\\1&\text{otherwise.}\end{array}\right .
\end{equation}
${\pmb k}$ and ${\pmb k'}$ denote pre-/post-flip states. A flip is always accepted if $A=1$; for $A<1$, it is accepted only if $A>r$ (with $r\in(0,1)$ uniformly random). As a local dynamics of Glauber type, Metropolis algorithm is suitable for studying nonequilibrium evolution~\cite{Metro-none,Metro-none-2}.

Testing a single spin is one Monte Carlo step. After $N$ steps (one sweep), all spins in the lattice have been flip-tested. Time $t$ is measured in sweeps.

The equilibration time $\tau_{\rm eq}$ is the number of sweeps needed for the order parameter to stabilize. Starting from an initial configuration, the magnetization evolves until reaching equilibrium, where it fluctuates around a mean value $\mu$ (with standard error $\sigma$). $\tau_{\rm eq}$ is recorded when the magnetization first enters the interval $(\mu-\sigma, \mu+\sigma)$.

According to the equilibrium statistical mechanics, the partition function in the canonical ensemble is
\begin{equation}
Z(T,H)=\sum_{\pmb{k}}\exp(-E_{\pmb{k}}/k_{\rm B}T).\label{partiton-f}
\end{equation}
The summation runs over all possible microstates. The Helmholtz free energy reads  
\begin{equation}
F(T,H)=-k_{\rm B}T\ln Z.\label{fp}
\end{equation}

To investigate the dependence of free energy on magnetization per spin, the restricted free energy method~\cite{1995-Lee,restricted-f} is widely employed in Monte Carlo simulations. It reads
\begin{equation}
F(m)=-k_{\rm B}T\ln \sum_{{\pmb{k}}}\delta(m_{{\pmb{k}}}-m)\exp(-E_{{\pmb{k}}}/k_{\rm B}T),\label{fm}
 \end{equation}
where $\delta$ is the Kronecker $\delta$ function. Namely, only those microstates whose magnetization per spin takes a given value are used to calculate the free energy.

In Monte Carlo calculations, when the system size is large, the value of the exponential function in Eq.~(\ref{fm}) overflows the variable's numerical range in programming. To overcome the problem, the energy per spin $\varepsilon_{\pmb{k}}=\frac{E_{\pmb{k}}}{N}$ is used instead of the total energy $E_{\pmb{k}}$ in Eq.~(\ref{fm}).

In this study, random initial configurations are employed based on our prior demonstration~\cite{XBLi2} that ultra-slow relaxation along first-order transition line emerges exclusively for random initial configurations. Three key features are: 
\begin{enumerate}[(1)]
\item{Exceptionally prolonged equilibration time - the mean value $\bar{\tau}_{\rm eq}$ exceeding those observed at the CP (see Fig.~2(a) in Ref.~\cite{XBLi2});}
\item{Critical-like finite-size scaling, i.e. $\bar{\tau}_{\rm eq}\sim L^z$, with larger exponent $z$ (see Fig.~2(b) in Ref.~\cite{XBLi2}). }
\item{Heavy-tailed distribution of $\tau_{\rm eq}$ (see Fig.~1(e) in Ref.~\cite{XBLi1}). }
\end{enumerate}

We investigate the time evolution of the free energy by sampling both the equilibrium-state configurations and the pre-equilibrium configurations. For pre-equilibrium configurations, we still employ the equilibrium free energy formula for calculation. For each temperature and evolution time, ten thousand configurations are generated. We bin ten thousand microstates into ten equal-width intervals based on their magnetization per spin ($m$), and calculate the free energy $F(m)$ in each bin.

\section{The free energy landscape and its time evolution along the phase boundary}

The free energy landscape $F(m)$ along the phase boundary is presented in Fig.~1. Our analysis focuses on six representative temperatures (4.60, 4.52, 4.51, 4.50, 4.49 and 4.20) under zero external field, covering the entire phase boundary. The time evolution is tracked at three evolution times ($t=2500$, $4500$ and $300,000$), represented by blue circles, green squares and red triangles, respectively. 

At $T=4.60$ ($T/T_{\rm c}=1.03$), all systems achieve thermal equilibrium by $t=2500$ (Fig.~1(h) in Ref.~\cite{XBLi1}). In contrast, for $T=4.20$ ($T/T_{\rm c}=0.93$), only a fraction of systems reach equilibrium by $t=2500$, with significant non-equilibrium populations persisting (Fig.~1(e) in Ref.~\cite{XBLi1}). The equilibrium fraction increases by $t=4500$, and almost all systems reach equilibrium at $t=300,000$. 


\begin{figure*}
     \centering
     \includegraphics[scale=0.35]{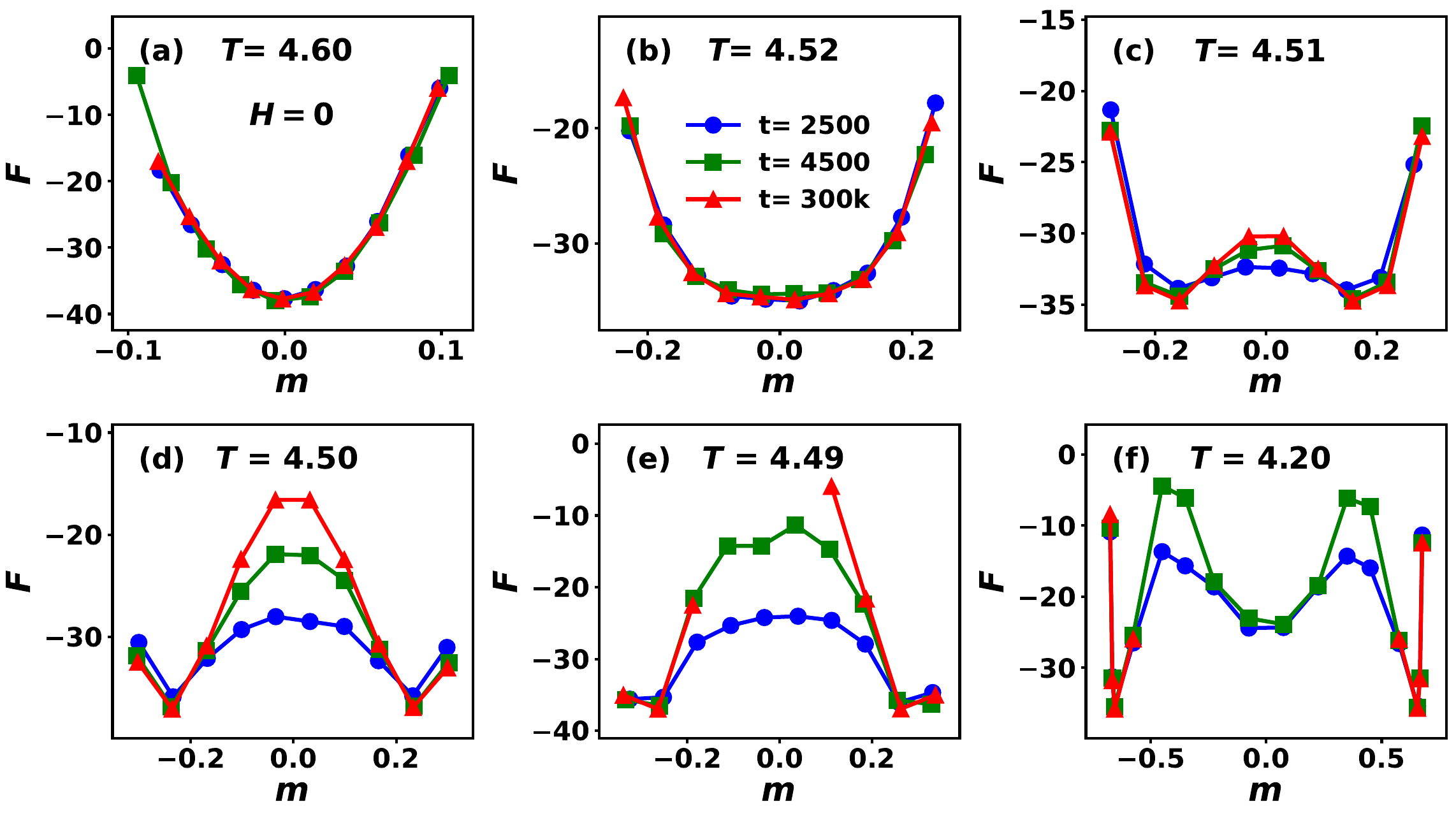}
      \caption{The free energy $F$ as a function of the magnetization per spin  for six values of temperature at $H=0$. The lattice size is $L$ = 60. }
\end{figure*}

Figure~1(a) shows the free energy $F(m)$ at $T=4.60$, displaying a V-shape that remains identical across all three time points. This temporal invariance reflects the extremely short relaxation time above $T_{\rm c}$, where equilibrium is rapidly achieved, making $F(m)$ time-independent. The single minimum at $m=0$ confirms the disordered phase as the stable equilibrium state.    

At $T=4.52$ (near $T_{\rm c}$) in Fig.~1(b), $F(m)$ shows a U-shape that is similarly time-independent, indicating that equilibrium is achieved before $t=2500$. Unlike Fig.~1(a), the minimum exhibits notable base widening - a characteristic of enhanced order parameter fluctuations near the CP.

At $T=4.51$ and $4.50$, Figs.~1(c-d) reveal the characteristic double-well structure predicted by Landau-Ginzburg theory, with degenerate minima corresponding to phase coexistence. The central maximum at $m=0$ represents the phase-separation barrier, whose height $\Delta F$ (quantified by the height difference between the maximum and the minimum) increases with evolution time while the overall landscape remains unchanged. The equilibrium ($t=300,000$) barrier heights are $\Delta F\approx 5$ ($T=4.51$) and $\approx22$ ($T=4.50$), demonstrating the temperature dependence predicted by $\tau^{\rm lifetime}\sim\exp(2\beta\sigma L^{d-1})$~\cite{1995-Lee}. This confirms that both the tunneling time and metastable state lifetime increase with decreasing temperature.

Due to finite-size effects, the critical temperature in our system is shifted upward from the thermodynamic limit value of $T_{\rm c}=4.51$. Consequently, at $T=4.51$, the system displays features of a 1st-PT rather than critical behavior.

Figure~1(e) shows the free energy at $T=4.49$ (slightly below $T_{\rm c}$). At shorter times ($t=2500$ and $4500$), $F(m)$ displays symmetric double minima at $m\approx\pm0.25$ separated by a finite barrier. By $t=300,000$, the central barrier region abruptly truncates, forming two disconnected wells. This reflects the system's evolution toward either positively-polarized or negatively-polarized states. The magnetization distribution becomes bimodal, with nearly all configurations clustered around the minima. Consequently, microstates with near-zero magnetization vanish entirely, leaving those bins devoid of statistics.

At $T=4.20$ (Fig.~1(f)), far below $T_{\rm c}$, the equilibrium states ($m\approx\pm0.65$) show stronger polarization. The free energy reveals a fine structure, i.e. an M-shape, between minima. To better capture the fine structure, the data have been further binned. The M-shaped barrier-well-barrier structure includes a metastable trapping potential well that deepens from $\Delta F\approx15$ to $25$ between $t=2500$ and $4500$. Ultimately at $t=300,000$, the landscape again shows barrier truncation, analogous to Fig.~1(e). 

In summary of this part, near the CP, our numerical results successfully reproduce the Landau-Ginzburg free energy structure. Beyond that, along the first-order transition line far from the critical point, we observe fine-structure features in the free energy landscape. Due to the presence of a metastable trapping potential well around $m=0$, the random initial configurations have to tunnel through the energy barrier to evolve toward either equilibrium state. This accounts for the ultra-slow relaxation of random initial states at low $T$, while polarized initial states relax rapidly.  

In order to further understand the relaxation dynamics near the phase boundary, Fig.~2 presents the free energy landscape for three external fields.

\begin{figure*}
     \centering
     \includegraphics[scale=0.21]{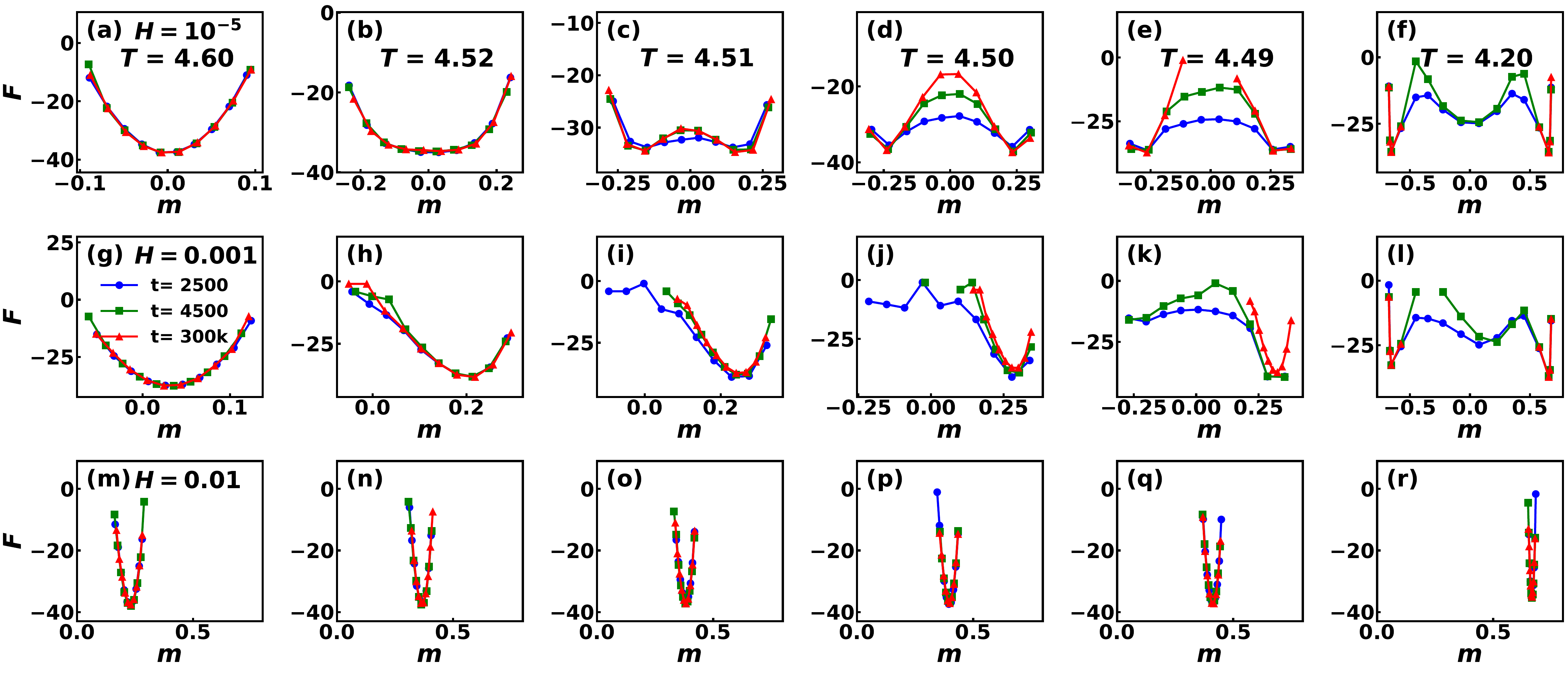}
      \caption{
      The free energy $F$ as a function of the magnetization per spin for six temperatures at three small external fields. Left to right: T=4.60, 4.52, 4.51, 4.50, 4.49, 4.20. Top to bottom: $H=10^{-5}$, $0.001$, $0.01$. The blue circles, green squares and red triangles correspond to three evolution times, $t=2500$, $4500$ and $300,000$, respectively. The lattice size is $L = 60$. }
\end{figure*}
 
Under extremely weak external field ($H=10^{-5}$), as depicted in Figs.~2(a-f), the free energy $F(m)$ maintains identical landscape to the zero-field case across all temperatures. When the field increases to $H=0.001$, notable modifications appear in the landscape. For temperatures $T=4.60$ and $4.52$ (Figs.~2g-h), the originally symmetric shape becomes asymmetric, with the minimum shifting rightward along the magnetization axis. The overlap of curves at different evolution times confirms that equilibrium is achieved before $t=2500$. The asymmetry reflects the field-induced symmetry breaking that favors magnetization alignment along the field direction. 

The Landau-Ginzburg theory predicts that the degenerate double-well structure becomes non-degenerate under a finite external field. The shallower well corresponds to a metastable state and the deeper well represents the true equilibrium state. This theoretical shape is qualitatively confirmed at $H=0.001$, where the blue circles in Figs.~2(i-k) shows the emergence of non-degenerate minima. However, as the system evolves to $t=300,000$, the free energy landscape undergoes a remarkable transformation - the metastable well disappears completely, leaving only a single, narrow minimum in the form of a skewed V-shape. This ultimate configuration reflects the system's inevitable progression toward absolute free energy minimization, where all transient metastable states must ultimately decay given sufficient evolution time.  

At $T=4.20$, the free energy landscape maintains its characteristic fine structure while developing asymmetry under applied field (Fig.~2l). The residual fine structure similarly impacts the relaxation dynamics, yielding a dynamic exponent $z=1.438\pm0.019$ - as much as one-third of the corresponding value ($z=3.763\pm0.057$) at $H=10^{-5}$ (Fig.~3 in Ref.~\cite{XBLi2}). The reduction in $z$ demonstrates how finite fields can accelerate the equilibration process in low-temperature systems. Upon reaching equilibration ($t=300,000$), the system evolves into two distinct potential wells of unequal depth. This broken symmetry results in preferential occupation of one polarized state over the other.

Figure~2(m-r) displays the free energy landscape under a stronger field ($H=0.01$). As predicted by Landau-Ginzburg theory, when the field exceeds a critical strength, the metastable minimum vanishes completely across all temperatures. In this regime, the free energy exhibits a simple sharp V-shape, with its minimum corresponding to an ordered phase aligned with the applied field direction. The complete overlap of the free energy landscape at different evolution times provides clear evidence of rapid equilibration under these conditions.

\section{Relative variance of the equilibration time and non-self-averaging}

The free energy structure on the first-order phase boundary far from $T_{\rm c}$ exhibits significant differences from $T_{\rm c}$, resulting in differences in equilibration times. The existence of coexistence states and metastable states at 1st-PT increases the randomness and uncertainty of the time evolution. This leads to an increase in the equilibration time, i.e. ultra-slow relaxation. Beyond that, the equilibration time exhibits significant fluctuations - ranging from short to exceptionally long - with a  distribution broader than that observed at $T_{\rm c}$, as addressed in our previous work~\cite{XBLi1}.

To comprehensively understand the relaxation dynamics, the relative variance of $\tau_{\rm eq}$ is defined as~\cite{nsa-cp-1}
\begin{equation}
    R_{\tau_{\rm eq}}=\frac{\overline{\tau_{\rm eq}^{2}}-\overline{\tau}_{\rm eq}^{2}}{\overline{\tau}_{\rm eq}^{2}},
\end{equation}
where bar represents an average over the sample. If $R_{\tau_{\rm eq}}$ tends to zero as the system size increases, $\tau_{\rm eq}$ is self-averaging. 

At the phase boundary, the relative variance of $\tau_{\rm eq}$ starting from random initial states is plotted against the system size for three different temperatures in Fig.~3(a). The trend of $R_{\tau_{\rm eq}}$ is obviously different for each temperature. At $T_{\rm c}=4.51$, $R_{\tau_{\rm eq}}$ remains almost constant as the system size increases. In other words, the variance of the $\tau_{\rm eq}$ distribution increases with the system size at the same rate as the square of the average of $\tau_{\rm eq}$, indicating non-self-averaging behavior of $\tau_{\rm eq}$. This trend is consistent with other observables~\cite{nsa-cp-1,nsa-cp-2,nsa-cp-3,nsa-cp-4}.  

At $T=4.30$ and $4.20$ which represent 1st-PT far from $T_{\rm c}$, it is shown that $R_{\tau_{\rm eq}}$ increases with system size, which we term \textit{self-diverging} behavior. This behavior is similar to what is observed in liquid crystals at low temperature~\cite{nsa-fo}. As the system size increases, the variance of $\tau_{\rm eq}$ increases more rapidly than the square of the average of $\tau_{\rm eq}$, and the distribution of $\tau_{\rm eq}$ become wider and flatter. This phenomenon suggests that randomness is significantly amplified with system size, leading to an abnormally increased variance. Self-diverging behavior is associated with such an extremely broad distribution of $\tau_{\rm eq}$, representing a previously unrecognized feature of ultra-slow relaxation phenomena of 1st-PT. 


\begin{figure}
\hspace{-0.3cm}
\includegraphics[scale=0.4]{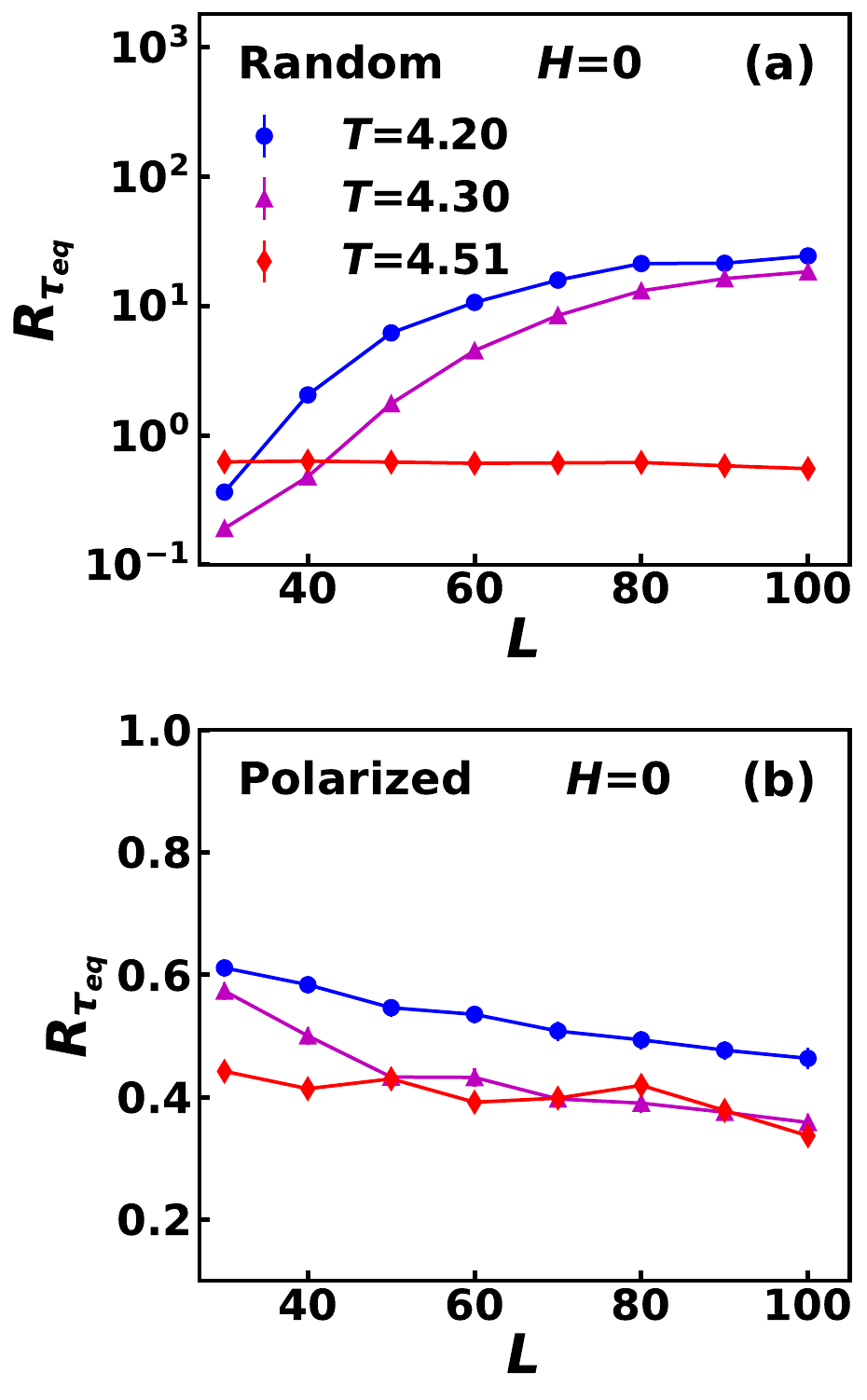}
\caption{The relative variance of $\tau_{\rm eq}$ as a function of system size for random initial states (a) and polarized initial states (b.) }
\end{figure}

For a reference, Fig.~3(b) presents the relative variance of $\tau_{\rm eq}$ starting from polarized initial states. It demonstrates that the relative variance of $\tau_{\rm eq}$ decreases with system size for all three temperatures, indicating self-averaging behavior of $\tau_{\rm eq}$ in this case. 

To investigate behavior beyond the phase boundary, Fig.~4 illustrates the relative variance $R_{\tau_{\rm eq}}$ at a given temperature $T = 4.20$, plotted against system size for three external fields. The external fields are $H = 10^{-5}$ (red circles), $H$ = 0.001 (blue triangles), and $H$ = 0.02 (green rhombuses).

It shows that self-diverging behavior is only prominent at $H = 10^{-5}$, which is the nearest to the first-order transition line. For stronger external fields, such as $H=0.001$ and $H=0.02$, $R_{\tau_{\rm eq}}$ decreases slowly as $L$ increases, indicating self-averaging of $\tau_{\rm eq}$. As the deviation from the first-order transition line becomes further, the self-diverging behavior of $\tau_{\rm eq}$  becomes less pronounced. Therefore, self-diverging behavior is observed not only precisely on the first-order transition line but also in its close vicinity, while far from the phase boundary, $\tau_{\rm eq}$ exhibits self-averaging behavior.

\begin{figure}
\includegraphics[scale=0.4]{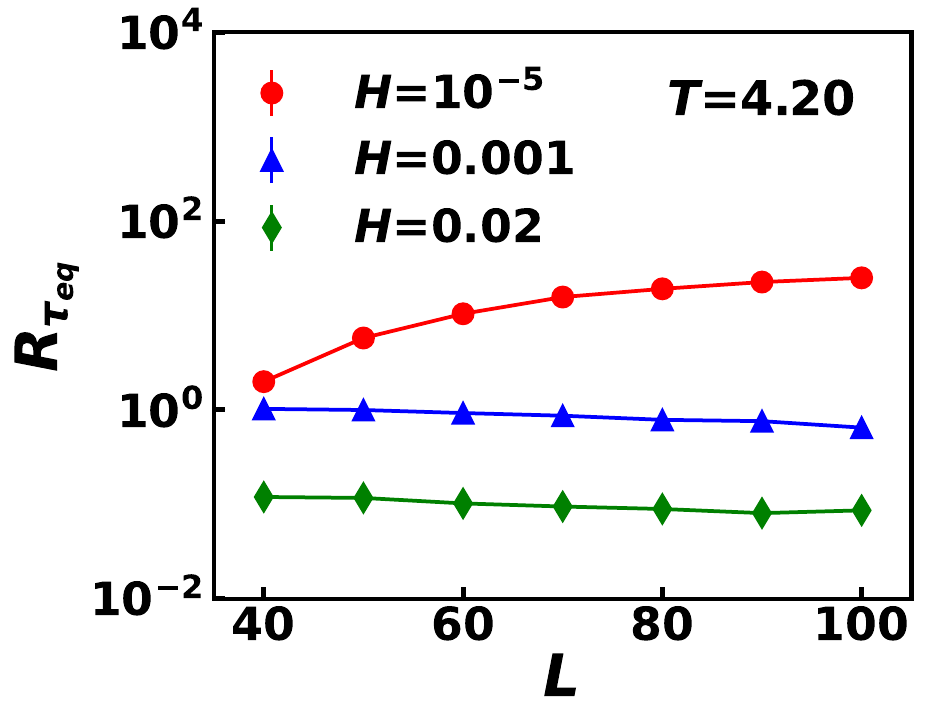}
\caption{The relative variance of $\tau_{\rm eq}$ as a function of system size for three external fields at $T = 4.20$. Evolution starts from random initial states.}
\end{figure}

\section{Summary and discussion}

In this work, we employ the Metropolis algorithm to simulate the relaxation of the three-dimensional Ising model from random initial configurations to equilibrium states along the entire phase boundary. The corresponding free-energy landscape is presented. Near the critical point, the characteristic double-well structure of the free energy is found to be consistent with predictions from Landau-Ginzburg theory.

For $T$ far below $T_{\rm c}$, the free energy landscape of pre-equilibrium configurations develops a barrier-well-barrier structure between the two stable equilibrium states, indicating a trapping potential well at $m=0$. The barrier becomes higher and higher when the system approaches equilibrium. Random initial configurations must tunnel through this barrier in order to evolve toward a stable equilibrium state. This results in observed ultra-slow relaxation.

Meanwhile, the relative variance of $\tau_{\rm eq}$ increases with system size, exhibiting a self-diverging behavior in contrast to non-self-averaging observed at the CP. This indicates an exceptionally broad distribution of equilibration times, revealing a previously unrecognized characteristic of the 1st-PT.

These two characteristics of the 1st-PT disappear gradually when the system is away from the phase boundary.

The ultra-slow relaxation and self-diverging behavior observed at the 1st-PT originate from the trapping potential well of the free energy landscape. On one hand, these features hinder the system from reaching equilibrium, making conventional equilibrium-based theories of phase transitions unreliable for identifying the phase boundary. On the other hand, these non-equilibrium characteristics themselves provide a distinct  signature for locating the phase boundary.


\vspace{0.5cm}
\section*{Acknowledgement}
This research was funded by the National Key Research and Development Program of China, Grant No. 2024YFA1610700, Grant No. 2022YFA1604900, and the National Natural Science Foundation of China, Grant No. 12275102. The numerical simulations have been performed on the GPU cluster in the Nuclear Science Computing Center at Central China Normal University (NSC3). 

\providecommand{\href}[2]{#2}\begingroup\raggedright\endgroup

\begin{thebibliography}{10}%
\makeatletter
\providecommand{\hrefCMSnoop }[0]{\@secondoftwo}%
\makeatother
\providecommand{\doi}{\texttt{doi:}\begingroup \urlstyle{tt}\Url}
\bibitem{1977}P. C. Hohenberg and B. I. Halperin, Rev. Mod. Phys. \textbf{49}, 435 (1977).
\bibitem{book1}E. M. Lifshitz and L. P. Pitaevskii, {\it Physical Kinetics} (Course of Theoretical Physics, Vol. 10) (reprinted by Beijing World Publishing Corporation by arrangement with Butterworth-Heinemann, 1999).
\bibitem{1979-Ikeda}H. Ikeda, Prog. Theor, Phys. \textbf{61}, 683 (1979).
\bibitem{2001-Joo}J. Joo, V. Oudovenko, Phys. Rev. B \textbf{64}, 193102 (2001).
\bibitem{2010-Fischer}T. Fischer, R. L. C. Vink, J. Phys.: Condens. Matter \textbf{22}, 104123 (2010).
\bibitem{1992-Berg}B. A. Berg, T. Neuhaus, Phys. Rev. Lett. \textbf{68}, 9 (1992).
\bibitem{1995-Lee}J. Lee, M. A. Novotny, P. A. Rikvold, Phys. Rev. E \textbf{52}, 356 (1995).
\bibitem{XBLi1}Xiaobing Li, Mingmei Xu, Yanhua Zhang, Zhiming Li, Yu Zhou, Jinghua Fu, Yuanfang Wu, Phys. Rev. C \textbf{105}, 064904 (2022).
\bibitem{XBLi2}Xiaobing Li, Ranran Guo, Mingmei Xu, Jinghua Fu,  Lizhu Chen, Yu Zhou,Yuanfang Wu, Phys. Rev. E \textbf{111}, 064115 (2025).
\bibitem{complex}K. Christensen and N. R. Moloney, {\it Complexity and Criticality} (Imperical College Press, London, 2005).
\bibitem{nsa-cp-1}S. Wiseman and E. Domany, Phys. Rev. E \textbf{52}, 3469 (1995).
\bibitem{nsa-cp-2}A. Malakis and N. G. Fytas, Phys. Rev. E \textbf{73}, 016109 (2006).
\bibitem{nsa-cp-3}G. Parisi and N. Sourlas, Phys. Rev. Lett. \textbf{89}, 257204 (2002).
\bibitem{nsa-cp-4}K. A. Pronin, Physica A \textbf{596}, 127180 (2022).
\bibitem{nsa-fo}J. M. Fish, R. L. C. Vink, Phys. Rev. Lett. \textbf{105}, 147801 (2010).
\bibitem{kc}A. L. Talapov and H. W. Blote, J. Phys. A: Math. Gen. \textbf{29}, 5727 (1996).
\bibitem{kinetic-Ising}N. Menyhard and G. Odor, Brazilian Journal of Physics \textbf{30}, 113 (2000).
\bibitem{Metro-none}C.-W. Liu, A. Polkovnikov and A. W. Sandvik, Phys. Rev. B \textbf{89}, 054307 (2014).
\bibitem{Metro-none-2}M. Acharyya, Phys. Rev. E \textbf{56}, 2407 (1997).
\bibitem{restricted-f}V. A. Abalmasov, SciPost Phys. \textbf{16}, 151 (2024).
\end{thebibliography}
\end{document}